\documentclass[aps,pra,amsfonts,amsmath,amssymb,showpacs]{revtex4}

\begin{document}

\title{Simon's Algorithm, Clebsch-Gordan  Sieves, and Hidden Symmetries of Multiple Squares}
\author{Dave Bacon}
\affiliation{Department of Computer Science \& Engineering and Department of Physics, University of Washington, Seattle, WA 98195}
\date{August 1, 2008}
\begin{abstract}
The first quantum algorithm to offer an exponential speedup (in the query complexity setting) over classical algorithms was Simon's algorithm for identifying a hidden exclusive-or mask.  Here we observe how part of Simon's algorithm can be interpreted as a Clebsch-Gordan transform.  Inspired by this we show how Clebsch-Gordan transforms can be used to efficiently find a hidden involution on the group $G^n$ where $G$ is the dihedral group of order eight (the group of symmetries of a square.)  This problem previously admitted an efficient quantum algorithm but a connection to Clebsch-Gordan transforms had not been made.  Our results provide further evidence for the usefulness of Clebsch-Gordan transform in quantum algorithm design.
\end{abstract} 
\maketitle

\section{Introduction}

The most widely known computational task under which quantum computers outperform classical computers is for the problem of factoring an integer~\cite{Shor:94a}.  This can be done in polynomial time on a quantum computer using Shor's algorithm, while the best classical algorithm requires superpolynomial time to complete this task.  Since the problem of factoring is the basis upon which the security of the most widely used public key cryptosystems is built~\cite{Rivest:78a}, Shor's algorithm is a threat to the security of our modern computer infrastructure.  While this is certainly of great importance, a question which has been burning a hole in the theoretical quantum computing community is what other problems can be exponentially sped up using a quantum computer.  Of particular significance are algorithms for the nonabelian hidden subgroup problem (defined below), as an efficient algorithm for these problems over the appropriate groups would lead to efficient algorithms for graph isomorphism~\cite{Boneh:95a,Beals:97a,Ettinger:99b,Hoyer:97a} and certain unique shortest vector in a lattice problems~\cite{Regev:02a}.

The first quantum algorithm which showed an exponential speedup (in the query complexity setting) over classical algorithms was the problem considered by Simon~\cite{Simon:94a}.  In Simon's problem, one is given access to a function on $n$ bits which is guaranteed to be constant and distinct on an unknown XOR mask and the goal of the problem is to identify this hidden XOR mask.  In the language of the hidden subgroup problem~\cite{Boneh:95a,Ettinger:99b}, Simon's problem is the hidden subgroup problem over the abelian group ${\mathbb Z}_2^n$ with the hidden subgroups being order two subgroups (involutions.)  Simon's algorithm consists of two components, one in which a Fourier transform over ${\mathbb Z}_2^n$ is applied, and second step in which the hidden subgroup is extracted from many samples of the first part and the use of Gaussian elimination.  Interestingly the first of these tasks follows naturally from symmetry considerations, while the second, at first glance, does not appear to follow from symmetry arguments.  Here we observe that,  upon further reflection, the Gaussian elimination in Simon's algorithm can indeed be interpreted in terms of symmetry.  In particular we observe that this step can be recast as a Clebsch-Gordan transform over ${\mathbb Z}_2^n$.  Recently Clebsch-Gordan transforms have emerged as a tool for solving some non-Abelian hidden subgroup problems~\cite{Kuperberg:03a,Alagic:06a,Bacon:06f}.  

A natural generalization of Simon's problem is to consider the hidden subgroup problem over $n$ copies of a constant sized finite non-Abelian group, i.e. the group ${\mathcal G}^n$, and consider the problem of hidden involutions for this group.  Recently Alagic, Moore, and Russell~\cite{Alagic:06a} have found subexponential, but still superpolynomial, time algorithms for this problem, under a certain representation theoretic restriction on the group ${\mathcal G}$.  Their algorithm relies on a method known as Clebsch-Gordan sieving, in which one uses a partial Clebsch-Gordan transform to sieve out particular irreps.  We have argued previously that beyond Clebsch-Gordan sieving one also needs to use information stored in the multiplicity register of a Clebsch-Gordan transform in a coherent fashion to solve hidden subgroup problems~\cite{Bacon:06f}.  Interestingly, there are groups, ${\mathcal G}$, for which the hidden subgroup problem is already known to admit efficient quantum algorithms (besides the obvious Abelian cases.)  For example, if ${\mathcal G}$ is  ${\mathcal D}_4$, the dihedral group of order eight, then the hidden involution problem over ${\mathcal D}_4^n$, then the algorithm of Friedl, Ivanyos, Magniez, Santha, and Sen~\cite{Friedl:03a} efficiently solves this problem (since ${\mathcal D}_4^n$ is a solvable group having a smoothly solvable commutator subgroups).  When applying the algorithm of Alagic, Moore, and Russell~\cite{Alagic:06a} to this group one obtains a superpolynomial time quantum algorithm for ${\mathcal D}_4^n$.  Is there a way to design a hidden subgroup problem algorithm which uses Clebsch-Gordan transforms, but not Clebsch-Gordan sieving, for this group?  Here we show how a quantum computer can efficiently solve such a hidden involution problem by using a multiplicity space algorithm for ${\mathcal D}_4^n$.  While we do not obtain a new efficient quantum algorithm, we are able to show how Clebsch-Gordan multiplicity space algorithms can be developed and lead to polynomial time quantum algorithms.  We present this as more evidence that multiplicity space Clebsch-Gordan transforms are a viable new tool for efficiently solving hidden subgroup problems.  

\section{The hidden subgroup problem}

The hidden subgroup problem has a long and vexing history in quantum computing (see \cite{Lomont:04a} for a review of this problem.  Note however that a considerable amount of progress has been made on this problem since this review was written.)  After Shor's discovery of efficient quantum algorithms for factoring and computing the discrete logarithm~\cite{Shor:94a}, it was quickly realized that these problems could be seen as instances of the hidden subgroup problem over abelian groups~\cite{Shor:94a,Kitaev:95a,Boneh:95a}.  The hidden subgroup problem is defined as follows:
\begin{quote}
{\bf Hidden Subgroup Problem.} (HSP)  Let $f$ be function from a group ${\mathcal G}$ to a set $S$ which is promised to be constant and distinct on different left cosets of an unknown subgroup ${\mathcal H}$: $f(g)=f(g^\prime)$ iff $g {\mathcal H}=g^\prime {\mathcal H}$.  The goal of the hidden subgroup problem is, by querying $f$, to identify the subgroup ${\mathcal H}$.  An algorithm for the hidden subgroup problem is efficient if the running time is polynomial in $\log |{\mathcal G}|$.
\end{quote}
The hidden subgroup problem can be efficiently solved in a variety of cases, including when the group is Abelian, the subgroups are normal, the group is extraspecial and more~\cite{Grigni:00a,Hallgren:00a,Ivanyos:01a,Friedl:03a,Moore:04a,Gavinsky:04a,Bacon:05a,Childs:05a,Ivanyos:07a,Ivanyos:07b}.  Two notable cases where there is no known efficient quantum algorithm for the hidden subgroup problem are the symmetric group hidden subgroup problem and the dihedral group hidden subgroup problem.  An efficient algorithm for the former would yield an efficient algorithm for the graph isomorphism problem~\cite{Boneh:95a,Beals:97a,Ettinger:99b,Hoyer:97a}, while an efficient algorithm for the later would yield an efficient algorithm for certain unique shortest vector in a lattice problems~\cite{Regev:02a}.  These two reductions form the foundation upon which interest in the hidden subgroup problem is based, since efficient algorithms for either of these two problems would expand the known power of quantum computers over classical computers for significant computational problems.

The standard approach to solving the hidden subgroup problem for the group ${\mathcal G}$ on a quantum computer is as follows.  First one creates a superposition over all possible group elements, ${1 \over \sqrt{|{\mathcal G}|}} \sum_{g \in {\mathcal G}} |g \rangle$. Following this one computes the function $f$ from the HSP in an ancillary register, producing the states
\begin{equation}
{1 \over \sqrt{|{\mathcal G}|}} \sum_{g \in {\mathcal G}} |g \rangle \otimes |f(g)]\rangle.
\end{equation}
Since the function $f$ is generally assumed to have no structure, this register is then measured or discarded.  If the function $f$ hides the subgroup ${\mathcal H}$, then this produces a random coset state $|g{\mathcal H}\rangle$, where
\begin{equation}
|g{\mathcal H}\rangle={1 \over \sqrt{|{\mathcal H}|}} \sum_{h \in {\mathcal H}} |gh\rangle.
\end{equation}
Here $g$ is a coset representative, and each of the coset states is produced at random with equal probability.  If we express this state as a density matrix, we obtain the so-called hidden subgroup state
\begin{equation}
\rho_{\mathcal H}={|{\mathcal H}| \over |{\mathcal G}|} \sum_{{\rm cosets~}g{\mathcal H}} |g {\mathcal H} \rangle \langle g{\mathcal H}|.
\end{equation}
The standard approach to the hidden subgroup problem attempts to construct efficient quantum circuits for identifying ${\mathcal H}$ given a polynomial number of copies of the hidden subgroup state $\rho_{\mathcal H}$.  

Given a hidden subgroup state $\rho_{\mathcal H}$ a symmetry argument immediately tells you that without a loss of generality, one can perform a unitary transform on this state which better expresses the structure of the information stored in $\rho_{\mathcal H}$.  In particular if we define the left regular representation of the group ${\mathcal G}$, via $D_L(g) |g^\prime\rangle =|gg^\prime\rangle$, then $\rho_{\mathcal H}$ commutes with this representation: $D_L(g) \rho_{\mathcal H}=\rho_{\mathcal H} D_L(g)$ for all $g \in {\mathcal G}$ and all possible subgroups ${\mathcal H}$.  Via Schur's lemma this tells us that there is a basis in which every $\rho_{\mathcal H}$ is block diagonal.  In particular we can decompose $\rho_{\mathcal H}$ into a direct sum of states as
\begin{equation}
\rho_{\mathcal H}=\bigoplus_{\mu} \sigma_{\mathcal H, \mu} \otimes I_{d_\mu},
\end{equation}
where the direct sum is over all irreps $\mu$ of $\mathcal G$, $d_{\mu}$ is the dimension of the $\mu$th irrep, and
\begin{equation}
\sigma_{\mathcal H,\mu} = {1 \over |{\mathcal G}|} \sum_{h \in {\mathcal H}} D_\mu(h),
\end{equation}
with $D_\mu$ is the $\mu$th irreducible representation (irrep) of $\mathcal G$.  The transform which block diagonalizes $\rho_{\mathcal H}$ is the quantum Fourier transform over $\mathcal G$.  We refer the reader to \cite{Bacon:06f} for details.  The main point here being, however, that the hidden subgroup state $\rho_{\mathcal H}$ is symmetric with respect to a representation of ${\mathcal G}$, and without loss of generality this symmetry implies that a unitary basis change can be made which better reveals the information stored in $\rho_{\mathcal H}$.  

\section{Simon's Algorithm and Clebsch-Gordan Transforms}

Simon's algorithm is the hidden subgroup problem on the group ${\mathbb Z}_2^n$ where the hidden subgroup is an order two subgroup.  We denote elements of this group as length $n$ bitstrings $z \in \{0,1\}^n$, and group multiplication simply corresponds to bitwise addition modulo $2$.  The hidden involution can be specified by a single bitstring $z \in \{0,1\}^n$ corresponding to the subgroup $\{0^n,z\}$.  We are guaranteed that the function $f$ hiding $z$ satisfies $f(x)=f(y)$ iff $y=x$ or $y=x+z$.  The representation theory of ${\mathbb Z}_2^n$ is quite simple.  Every irrep is one dimensional and is parameterized by a vector $r \in \{0,1\}^n$.  In particular the irrep is given by $D_r(x)=(-1)^{x\cdot r}$, where $a \cdot b = \sum_{i=1}^n a_i b_i ~{\rm mod}~2$.   

Let us briefly review Simon's original algorithm for this problem~\cite{Simon:94a}.  Simon's algorithm proceeds by the standard method for hidden subgroup algorithms on quantum computers.  Following our description of the standard method above, we see that this produces the random coset state 
\begin{equation}
{1 \over \sqrt{2}}(|x\rangle+|x+z\rangle),
\end{equation}
where $x$ is chosen uniformly at random from $\{0,1\}^n$.   Following our discussion above, one can then, without loss of generality performs a Fourier transform over ${\mathbb Z}_2^n$ on this hidden subgroup state.  The Fourier transform over ${\mathbb Z}_2^n$ is nothing more than $H^{\otimes n}$ where $H$ is the Hadamard transform,
\begin{equation}
H={1 \over \sqrt{2}}\left[\begin{array}{cc} 1 & 1 \\ 1 & -1 \end{array} \right].
\end{equation}
If one applies $H^{\otimes n}$ on the random coset state one obtains the state
\begin{equation}
{1 \over \sqrt{2^{n+1}}} \sum_{y \in \{0,1\}^n} \left[ (-1)^{x \cdot y } + (-1)^{x \cdot (y+z)} \right] |y\rangle ={1 \over \sqrt{2^{n+1}}} \sum_{y \in \{0,1\}^n} (-1)^{x \cdot y } \left[1+ (-1)^{x \cdot z} \right] |y\rangle. 
\end{equation}
Next notice that if you measure this state you will obtain a random vector $y$ such that $y \cdot z=0$.  Notice, importantly, here that in terms of group representation theory, $y \in \{0,1\}^n$, is an irrep label.  Thus we perform the quantum Fourier transform over ${\mathbb Z}_2^n$ and measure the irrep label.  Up to this stage, all of Simon's algorithm could have been motivated by simple observations about the symmetry of the hidden subgroup state.  But the next stage of the algorithm is does not appear to have a symmetry argument.  In particular, the next step in Simon's algorithm is to perform the above procedure $n-1$ times and then uses Gaussian elimination to identify $y$.  Note that with high probability random $y$'s such that $y \cdot z=0$ are linearly independent, so that this procedure succeeds with high probability.  While the Gaussian elimination here is an obvious approach to the problem, a natural question to ask is whether this part of the transform can be interpreted in terms of symmetry.

Consider two irreducible representations of a group ${\mathcal G}$: $D_{\mu_1}$ and $D_{\mu_2}$.  Then there is a representation of this group, which is called the direct product representation, given by $D(g)=D_{\mu_1}(g) \otimes D_{\mu_2}(g)$.  Since this is a representation of the group, it is decomposable into a direct sum of irreducible representations of ${\mathcal G}$:
\begin{equation}
D(g)=D_{\mu_1}(g) \otimes D_{\mu_2}(g)= \bigoplus_\mu I_{n_{\mu_1,\mu_2}^\mu}  \otimes D_{\mu}(g),
\end{equation}
where $n_{\mu_1,\mu_2}^\mu$ is the number of times irrep $\mu$ appears in this representation.  The unitary transform that enacts the above basis change is the Clebsch-Gordan transform~\cite{Sakari:94a,Chen:02a,Bacon:06f}. 

For the group ${\mathbb Z}_2^n$, like for all Abelian groups, the Clebsch-Gordan transform is rather simple.  Recall that irreps of ${\mathbb Z}_2^n$ are parameterized by vectors $r \in \{0,1\}^n$.  Then the Clebsch-Gordan transform is
\begin{equation}
D_{r_1}(g) \otimes D_{r_2}(g) = D_{r_1+r_2}(g),
\end{equation}  
where $r_i$ are the irrep labels and $r_1+r_2$ is the new irrep label produced by bitwise addition modulo $2$.  In other words, a Clebsch-Gordan transform over ${\mathbb Z}_2^n$ corresponds to nothing more than bitwise addition modulo $2$ of the irrep labels.  But this is exactly what is performed in Gaussian elimination: one selectively performs addition between the different $y$ vectors, which, recall, are irrep labels.  In other words, the Gaussian elimination step in Simon's algorithm can be reinterpreted as selective Clebsch-Gordan transforms over ${\mathbb Z}_2^n$.  Thus while this last step of Simon's algorithm is usually not understood in terms of a representation theoretic explanation, one can indeed provide such an interpretation by noting that this step is nothing more than a Clebsch-Gordan sieve\cite{Kuperberg:03a,Alagic:06a,Bacon:06f}.

Given that there is a Clebsch-Gordan transform hidden inside of Simon's algorithm, a natural question is to ask whether Clebsch-Gordan transforms can be used to efficiently solve the hidden involution problem for groups ${\mathcal G}^n$ for some constant sized group ${\mathcal G}$.  Further, as has been argued previously, an important class of algorithms are those which use the multiplicity space of the Clebsch-Gordan transform \cite{Bacon:06f}.  Here we will consider the hidden involution problem for the group ${\mathcal D}_4^n$ where ${\mathcal D}_4$ is the dihedral group of order eight.  This group is smoothly solvable (it is solvable, and has abelian factor groups of constant exponent) and therefore the algorithm of Friedl, Ivanyos, Magniez, Santha, and Sen~\cite{Friedl:03a} can be used to solve the hidden subgroup problem over this group.  We return to this group thus not to derive a new hidden subgroup algorithm, but to explore a small example of a hidden subgroup problem where one can make progress by thinking about Clebsch-Gordan transforms.  This complements our previous work wherein we showed that Clebsch-Gordan transforms could be used to understand how an efficient quantum algorithm for the Heisenberg hidden subgroup problem works~\cite{Bacon:05a,Bacon:06f}.

\section{The Group ${\mathcal D}_4$}

We begin by describing some relevant facts about the group ${\mathcal D}_4$. ${\mathcal D}_4$ is the group of symmetries of a square (it is not the group of the quaternions which is the other non-abelian group of order eight.)  It has eight group elements which we will label by $r^t s^k$ where $t \in \{0,1\}$ and $s \in \{0,1,2,3\}$ and is defined by the multiplication rule $r^t s^k r^{t^\prime} s^{k^\prime}=r^{t+t^\prime} s^{(-1)^t k +k^\prime}$.  ${\mathcal D}_4$ is a semidirect product of ${\mathbb Z}_4$ and ${\mathbb Z}_2$.

The representation theory of ${\mathcal D}_4$ is straightforward~\cite{Serre:77a}.  There are five different irreducible representations (irreps) of ${\mathcal D}_4$, four one dimensional irreps and one two dimensional irrep.  The one dimensional irreps are given by
\begin{eqnarray}
D_{t}(r^ts^k)=1, \quad D_{a}(r^ts^k)=(-1)^t, \quad D_{r}(r^ts^k)=(-1)^k, \quad {\rm and} \quad D_{ra}(r^ts^k)=(-1)^{k+t},
\end{eqnarray}
while the two dimensional irrep is given by
\begin{equation}
D_{1}(r^t s^k)=\left[ \begin{array}{cc} \omega^k \delta_{t,0} & \omega^{-k} \delta_{t,1} \\
\omega^{k} \delta_{t,1} & \omega^{-k} \delta_{t,0}
 \end{array}  \right],
\end{equation}
where $\omega=\exp \left[ {2 \pi i \over 4} \right]=i$.  Instead of using these irreducible representations, we will find it useful to introduce the follow two dimensional (sometimes reducible) representations of ${\mathcal D}_4$,
\begin{equation}
D_{j}(r^t s^k)=\left[ \begin{array}{cc} \omega^{jk} \delta_{t,0} & \omega^{-jk} \delta_{t,1} \\
\omega^{jk} \delta_{t,1} & \omega^{-jk} \delta_{t,0}
 \end{array}  \right], \label{eq:rep}
\end{equation}
where $j \in {\mathbb Z}_4$.  It is easy to check that $D_0$ is reducible to $D_a$ and $D_t$, $D_2$ is reducible to $D_r$ and $D_{ra}$, and $D_3$ is equivalent to $D_1$.  Also note that $X D_j X = D_{-j}$ where $X$ is the Pauli $X$ operator. 

There are ten different subgroups of ${\mathcal D}_4$.  We will be interested in the trivial and order two subgroups which contain a reflection.  We let ${\mathcal H}_{0,0}=\{e\}$ be the trivial subgroup and ${\mathcal H}_{1,l}=\{e,rs^l\}$, $l \in \{0,1,2,3\}$ be the order two subgroups labeled naturally.  Note that ${\mathcal H}_{1,0}$ is conjugate to ${\mathcal H}_{1,2}$ and ${\mathcal H}_{1,1}$ is conjugate to ${\mathcal H}_{1,3}$.

The motivation behind our algorithm is the Clebsch-Gordan series for ${\mathcal D}_4$.  In fact we will be most interested in the Clebsch-Gordan series, not over the irreps, but over the (sometimes) reducible representations $D_j$ defined in Eq.~(\ref{eq:rep}).  In this case the Clebsch-Gordan series is
\begin{equation}
D_i(g) \otimes D_j(g) = D_{i+j}(g) \oplus D_{i-j}(g),
\end{equation}
where $i,j \in {\mathbb Z}_4$ and the addition and subtraction are done mod $4$.

\section{Simon's Problem on ${\mathcal D}_4^n$.}

We will consider Simon's problem on ${\mathcal D}_4^n$.  In particular we will consider the HSP over ${\mathcal D}_4^n$ when the hidden subgroup is order two.  We can label these subgroups by
\begin{equation}
((t_1,l_1),(t_2,l_2),\dots,(t_n,l_n)),\quad {\rm where}\quad (t_i,l_i) \in \{ (0,0),(1,0),(1,1),(1,2),(1,3)\},
\end{equation}
corresponding to the subgroup consisting of the identity and the element $ \times_{i=1}^n r^{t_i} s^{k_i}$ (if all the $t_i=0$ only the identity element is included, we will exclude this case for now.)  It will be useful to label the hidden subgroup by the vectors $t \in {\mathbb Z}_2^n$ and $l \in {\mathbb Z}_4^n$.

We will follow the standard method for the hidden subgroup problem.  When we query the function $f$ which hides the hidden subgroup ${\mathcal H}$ in superposition over all possible group elements we obtain the state
\begin{equation}
{1 \over \sqrt{8^n}} \sum_{t \in {\mathbb Z}_2^n} \sum_{k \in {\mathbb Z}_4^n} |\times_{i=1}^n r^{t_i} s^{k_i} \rangle \otimes |f(\times_{i=1}^n r^{t_i} s^{k_i}) \rangle.
\end{equation}
If we now throw away the register where the function has been evaluated, we obtain a random coset state
\begin{eqnarray}
|u,k\rangle&=&{1 \over \sqrt{2}} \left( |\times_{i=1}^n r^{u_i} s^{k_i} \rangle + |\times_{i=1}^n (r^{u_i} s^{k_i} r^{t_i} s^{l_i} \rangle \right) \nonumber \\
&=&{1 \over \sqrt{2}} \left( |\times_{i=1}^n r^{u_i} s^{k_i} \rangle + |\times_{i=1}^n (r^{u_i+t_i}  s^{(-1)^{t_i} k_i+l_i} \rangle \right),
\end{eqnarray}
where $u \in {\mathbb Z}_2^n$ and $k \in {\mathbb Z}_4^n$.  We will obtain a particular coset state with uniform equal probability over all possible $u$ and $k$ vectors.  Now suppose that we perform a quantum Fourier transform or its inverse over ${\mathbb Z}_4$ on the individual $s$ registers conditional on whether the $r$ register is $e$ (forward QFT) or $r$ (inverse QFT) for the above state.  If we then measure the resulting register we will obtain $\mu \in {\mathbb Z}_4^n$, uniformly at random, and the state
\begin{equation}
{1 \over \sqrt{2}}\left(|\times r^{u_i} \rangle + \omega^{\sum_{i=1}^n (-1)^{u_i+t_i} \mu_i l_i} |\times r^{u_i+t_i}\rangle \right).
\end{equation}
If instead of using the group elements to label these states, we use a binary bit string to represent this register, we obtain the state
\begin{equation}
{1 \over \sqrt{2}} \left( |b\rangle + \omega^{\sum_{i=1}^n (-1)^{b_i+t_i} \mu_i l_i} |b+t\rangle \right), \label{eq:coset}
\end{equation}
where $b \in {\mathbb Z}_2^n$ and the addition $b+t$ is done componentwise.  Thus to recap, the above procedure produces a uniformly random $\mu \in {\mathbb Z}_4^n$ along with the state in Eq.~(\ref{eq:coset}) where $b$ is chosen uniformly from ${\mathbb Z}_2^n$.  Recall that we wish to determine $t$ and $l$.

\section{Clebsch-Gordan Transform Motivation for the Quantum Algorithm}

Now we will explain in slightly more detail our motivation for the algorithm which we are about to derive.  A variation on the hidden subgroup problem which is often as difficult as the full hidden subgroup problem is to identify, instead of the subgroup, the set of the conjugate subgroups to which the hidden subgroup belongs.  This problem is called the hidden subgroup conjugacy problem~\cite{Bacon:06f}.  Two subgroups ${\mathcal H}_1 \subset {\mathcal G}$ and ${\mathcal H}_2 \subset {\mathcal G}$ are conjugate to each other if there exists an element of $g \in {\mathcal G}$, such that ${\mathcal H}_1=\{ghg^{-1}, h \in {\mathcal H}_2\}$.  The notion of conjugate subgroups forms an equivalence relation among subgroups.  For many groups, including the important symmetric group, the hidden subgroup conjugacy problem is equivalent to the hidden subgroup problem~\cite{Fenner:06a,Bacon:06f}.  The hidden subgroup conjugacy problem, when cast as a state identification problem for quantum states, has an extra symmetry.  In particular for the case where we have queried a the hidden subgroup $m$ times using the standard method, the state distinction problem has a symmetry related to the diagonal action of the group on these states.  We refer the reader to~\cite{Bacon:06f} for a more detailed discussion of this symmetry.

In \cite{Bacon:06f} the symmetry of the hidden subgroup conjugacy problem states was shown to lead naturally to the Clebsch-Gordan transform over the relevant finite group.  In \cite{Bacon:06f} it was shown that for the hidden subgroup conjugacy problem, information about the hidden subgroup conjugacy is found in the multiplicity space of multiple copies of the hidden subgroup states.  For the case of the Heisenberg hidden subgroup group, a case which had previously been shown to admit an efficient quantum algorithm in~\cite{Bacon:05a}, it was shown that measurement of the multiplicity space could be used to solve the hidden subgroup problem efficient~\cite{Bacon:06f}.  Thus motivated, we can examine the hidden subgroup conjugacy problem for $n$ copies of the dihedral group where the hidden subgroups are involutions.  For the dihedral group of order eight, the subgroups, $\{e,r\}$ and $\{e,rs^s\}$ are conjugate to each other, as are $\{e,rs\}$ and $\{e,rs^3\}$.  For the reducible $D_i$ representations described above, the Clebsch-Gordan transform is rather simple, being related to the simple Clebsch-Gordan series $D_i(g) \otimes D_j(g)=D_{i+j}(g) \oplus D_{i-j}(g)$.  Define the double controlled-not
\begin{equation}
U=\left[
    \begin{array}{cccc}
      1 & 0 & 0 & 0 \\
      0 & 0 & 0 & 1 \\
      0 & 1 & 0 & 0 \\
      0 & 0 & 1 & 0 \\
    \end{array}
  \right]
\end{equation}
This operation takes the computational basis state $|x,y\rangle$ to $|x\oplus y, x\rangle$.   If we apply this transform to two reducible representation $D_i$ and $D_j$, this transform enacts a Clebsch-Gordan transform.  In particular,
\begin{equation}
U (D_i (g) \otimes D_j(g)) U^\dagger= \bigoplus_{s=\pm 1} D_{i +s j}(g).
\end{equation}

In other words, after applying the $U$ transform, the first qubit will contain the multiplicity of the new representation and the second qubit will be the space where this representation acts.  For ${\mathcal D}_4^n$, a bitwise application of $n$ $U$s is the Clebsch-Gordan transform over ${\mathcal D_4^n}$.  Now from analysis of the hidden subgroup conjugacy problem we know that information about the subgroup conjugacy of the hidden subgroup must lie in the multiplicity space after a Clebsch-Gordan transform.  Thus motivated our algorithm will proceed from exactly this first step to produce an efficient quantum algorithm.

\section{The Quantum Algorithm}

Recall from the previous section that the standard method plus a conditional quantum Fourier transform produces a uniformly random $\mu \in {\mathbb Z}_4^n$ and $b \in {\mathbb Z}_2^n$ along with the state
\begin{equation}
{1 \over \sqrt{2}} \left( |b\rangle + \omega^{\sum_{i=1}^n (-1)^{b_i+t_i} \mu_i l_i} |b+t\rangle \right) \label{eq:bsup}
\end{equation}
Our quantum algorithm will proceed in three stages.  In the first state we will determine $t$.  In the second state we will determine the parity of $l$.  In the final stage will determine $l$.

\subsection{Determining $t$}

Suppose we have two coset states
\begin{eqnarray}
{1 \over \sqrt{2}} \left( |b_1\rangle + \omega^{\sum_{i=1}^n (-1)^{(b_1)_i+t_i} (\mu_1)_i l_i} |b_1+t\rangle \right) \otimes {1 \over \sqrt{2}} \left( |b_2\rangle + \omega^{\sum_{i=1}^n (-1)^{(b_2)_i+t_i} (\mu_2)_i l_i} |b_2+t\rangle \right),
\end{eqnarray}
and we apply, bitwise, $n$ double controlled-not's ($U$s) between these two registers and the addition is done over ${\mathbb Z}_2^n$.  This will produce the state
\begin{eqnarray}
&&{1 \over 2}|b_1 + b_2\rangle \otimes \left( |b_1\rangle+\omega^{\sum_{i=1}^n [(-1)^{(b_1)_i+t_i} (\mu_1)_i +(-1)^{(b_2)_i+t_i} (\mu_2)_i] l_i  }|b_1+t\rangle \right) \nonumber \\
&&+{1 \over 2}|b_1 + b_2 +t\rangle \otimes \left(  \omega^{\sum_{i=1}^n (-1)^{(b_2)_i+t_i} (\mu_2)_i l_i}|b_1\rangle+  \omega^{\sum_{i=1}^n (-1)^{(b_1)_i+t_i} (\mu_1)_i l_i}|b_1+t\rangle \right).
\end{eqnarray}
We can rewrite this in the form
\begin{eqnarray}
&&{1 \over 2}|b_1 + b_2\rangle \otimes \left( |b_1\rangle+\omega^{\sum_{i=1}^n (-1)^{(b_1)_i+t_i}[ (\mu_1)_i +(-1)^{(b_1)_i+(b_2)_i} (\mu_2)_i] l_i  }|b_1+t\rangle \right) \nonumber \\
&&+{1 \over 2}\omega^{\sum_{i=1}^n (-1)^{(b_2)_i+t_i} (\mu_2)_i l_i}|b_1 + b_2 +t\rangle \otimes \left(  |b_1\rangle+  \omega^{\sum_{i=1}^n (-1)^{(b_1)_i+t_i}[ (\mu_1)_i -(-1)^{(b_1)_i+(b_2)_i} (\mu_2)_i] l_i  }|b_1+t\rangle \right).
\end{eqnarray}
A further simplification is to write this in a summed form:
\begin{equation}
{1 \over 2}\sum_{c_2 \in {\mathbb Z}_2} \omega^{\sum_{i=1}^n (-1)^{(b_2)_i+t_i } (\mu_2)_i l_i c_2}|b_1 + b_2 +c_2 t\rangle \otimes \left(  |b_1\rangle+  \omega^{\sum_{i=1}^n (-1)^{(b_1)_i+t_i}[ (\mu_1)_i +(-1)^{(b_1)_i+(b_2)_i+c_2} (\mu_2)_i] l_i  }|b_1+t\rangle \right).
\end{equation}
Notice, now that the second register contains a state just like the original coset states, except now for different $\mu$'s.  In particular, for the first part of the superposition, the new $\bar{\mu}$ has $\bar{\mu}_i=(\mu_1)_i + (-1)^{(b_1)_i+(b_2)_i} (\mu_2)_i$, while for the second part of the superposition, the new $\bar{\mu}$ has $\bar{\mu}_i=(\mu_1)_i - (-1)^{(b_1)_i+(b_2)_i} (\mu_2)_i$.  Note that the sign of this addition of $\mu_1$ and $\mu_2$ components is given by the bitstring in the first register.  Indeed, the above transform corresponds exactly to the Clebsch-Gordan transform over ${\mathcal D}_4^n$.

Suppose that we produce $m$ coset states and in a cascade perform the double control-not operation on these states.  The resulting state will be
\begin{eqnarray}
{1 \over \sqrt{2^m}}\sum_{c_2,\dots,c_m \in {\mathbb Z}_2} &&\omega^{\sum_{j=2}^m \sum_{i=1}^n (-1)^{(b_j)_i+t_i} (\mu_j)_i l_i c_j}
 |b_1+b_2+c_2 t\rangle \otimes |b_1+b_3+c_3 t\rangle \otimes \cdots \otimes |b_1+b_m+c_m t\rangle \nonumber \\
&& \otimes \left(  |b_1\rangle+  \omega^{\sum_{i=1}^n (-1)^{(b_1)_i+t_i}[ (\mu_1)_i +\sum_{j=2}^m (-1)^{(b_1)_i+(b_j)_i+c_j } (\mu_j)_i] l_i  }|b_1+t\rangle \right)
\end{eqnarray}
Now notice the following.  From the first $m-1$ registers, along with the values of the $\mu_j$ we can compute part of that phase which appears in the last register
\begin{equation}
(\mu_{tot})_i= (\mu_1)_i +\sum_{j=2}^m (-1)^{(b_1)_i+(b_j)_i+c_j t_i} (\mu_j)_i. \label{eq:phase}
\end{equation}
We can compute this value an place it in an ancilla register, thus producing the state
\begin{eqnarray}
&&{1 \over \sqrt{2^m}}\sum_{c_2,\dots,c_m \in {\mathbb Z}_2} \omega^{\sum_{j=2}^m \sum_{i=1}^n (-1)^{(b_j)_i+t_i} (\mu_j)_i l_i c_j}
 |b_1+b_2+c_2 t\rangle \otimes |b_1+b_3+c_3 t\rangle \otimes \cdots \otimes |b_1+b_m+c_m t\rangle \\
&& \otimes \left(  |b_1\rangle+  \omega^{\sum_{i=1}^n (-1)^{(b_1)_i+t_i}[ (\mu_1)_i +\sum_{j=2}^m (-1)^{(b_1)_i+(b_j)_i+c_j } (\mu_j)_i] l_i  }|b_1+t\rangle \right) \otimes \bigotimes_{i=1}^n|(\mu_1)_i +\sum_{j=2}^m (-1)^{(b_1)_i+(b_j)_i+c_j t_i} (\mu_j)_i \rangle. \nonumber 
\end{eqnarray}
We will now measure this last ancilla register.  Now for $m$ much less than $n$, the most likely outcome is that the above superposition will contain a unique value in this last register for each term in the superposition (ignoring superpositions over the $m$th register.)  This is because for random $\mu_i$ the phase in Eq.~(\ref{eq:phase}) a random $\pm$ combination of these terms will likely only produce only one solution.  This changes, however when $m=n+2$.

In order to understand this, consider the following problem, motivated by the hidden subgroup problem over ${\mathcal D}_4$ (which is not hard, but which is illustrative.) In this problem you are given $k$ random $v_i \in {\mathbb Z}_4$.  Consider the function from $\{+1,-1\}^k$ to $\{0,1,2,3\}$,
\begin{equation}
f(s_1,\dots,s_k)=\sum_i s_i v_i {\rm ~mod~} 4.
\end{equation}
First note that because $-0=0~{\rm mod}~4$, $-2=2~{\rm mod}~4$, $-1=3~{\rm mod}~4$, and $-3=1~{\rm mod}~4$, the parity of $f(s_1,\dots,s_k)$ for all $s_i \in \{+1,-1\}$ is the same.  In other words, for any possible $s_i = \pm 1$, $f$ can either be in $\{0,2\}$ or $\{1,3\}$.  Thus suppose you are given random $v_i \in {\mathbb Z}_4$, and a value of $v_{tot}=f(s_1,\dots,s_k)$ for some fixed $s_i$.  Given this information, you are then required to return a list of $s_i$ whose sign can be flipped and still retain the total $v_{tot}$.  This can easily be done as follows.  Let $w_i$ denote the parity of $v_i$ (i.e. $0$ if $v_i \in \{0,2\}$ and $1$ otherwise.)  If $w_i=0$, then flipping the sign of $s_i$ will produce the same $v_{tot}$, since $\pm 0 =0$ and $\pm 2=2$ over ${\mathbb Z}_4$.  Further since $\{1,3\} \pm \{1,3\} \in \{0,2\}$, flipping any even number of $s_i$ where $w_i=1$ also keeps the $v_{tot}$.  In other words if you consider all vectors $y \in {\mathbb Z}_2^k$ such that $y \cdot w =0$, then simultaneously flipping the sign where $y_i=1$ preserves the sum.  (Note there is a degenerate case where all $w_i=0$ which we will ignore for now.)

Now extend this problem to the case we are considering where $m=n+2$.  For each $i$ we can apply the above procedure to the numbers $(\mu_2)_i,\dots,(\mu_m)_i$ with a total of $(\mu_{tot})_i-(\mu_1)_i$.  This will produce a $w_i$ such that simultaneously flipping the $w_i=1$ bits where $w_i \cdot w_i =0$ will keep this term in the sum the same.  Since we need to simultaneously flip the bits not just for a fixed $i$ but across all $i$ and retain the same total, we will obtain a series of $n$ equation $w_i \cdot s=0$, where the $w_i \in {\mathbb Z}_2^{m-1}={\mathbb Z}_2^{n+1}$.  With probability greater than one half we will obtain a $s$ which satisfies these equations (and this vector can be found efficiently using Gaussian elimination.)

Thus with high probability we have identified a vector $s \in {\mathbb Z}_2^{n+1}$ indicating which of the $m=2,\dots n+2$ signs can be flipped to obtain the same sum in the last register.  Having identified those locations where the sign can be flipped an obtain the same sum, now consider the $i$th registers where $s_i=1$.  The state of these registers will be
\begin{equation}
{1 \over \sqrt{2}} \left[\omega^{\sum_{j|s_j=1}^m \sum_{i=1}^n (-1)^{(b_j)_i+t_i} (\mu_j)_i l_i c_j} \bigotimes_{i|s_i=1} |b_1+b_i+c_i t \rangle+\omega^{\sum_{j|s_j=1}^m \sum_{i=1}^n (-1)^{(b_j)_i+t_i} (\mu_j)_i l_i (1-c_j)}  \bigotimes_{i|s_i=1} |b_1+b_i+(1-c_i) t \rangle \right]
\end{equation}
where $c_j$ are now some fixed but unknown $\{0,1\}$s.  The global phase can be pulled out obtaining
\begin{equation}
{1 \over \sqrt{2}} \left[ \bigotimes_{i|s_i=1} |b_1+b_i+c_i t \rangle+\omega^{\sum_{j|s_j=1}^m \sum_{i=1}^n (-1)^{(b_j)_i+t_i} (\mu_j)_i l_i (1-2c_j) }  \bigotimes_{i|s_i=1} |b_1+b_i+(1-c_i) t \rangle \right]
\end{equation}
Note that by measuring the other registers where $s_i=0$  and using the value of the total $\mu_{tot}$ in the ancilla register, we can calculate in another ancilla register the value
\begin{equation}
\sum_{j=2|s_j=1}^m (-1)^{(b_1)_i+(b_j)_i+c_j t_i} (\mu_j)_i
\end{equation}
A consequence of finding the $s_i=1$ where flipping the sign of the $\pm \mu_i$ terms obtain the same sum is that the $\pm \mu_i$ sums over these $i$ must all be made up of vectors from $\{0,2\}^n$.  Thus the above terms must be in $\{0,2\}$.  Since $\pm 0 =0$ and $\pm 2=2$ over ${\mathbb Z}_4$, this sum must be
\begin{equation}
r_i=-\sum_{j=2|s_j=1}^m (-1)^{(b_j)_i+c_j t_i} (\mu_j)_i. \label{eq:02}
\end{equation}
Now noting that the phase in Eq.~(\ref{eq:02}) is zero where $t_i=0$ (since this implies $l_i=0$), we can express the state in Eq.~(\ref{eq:02}) as
\begin{equation}
{1 \over \sqrt{2}} \left[ \bigotimes_{i|s_i=1} |b_1+b_i+c_i t \rangle+\omega^{\sum_{i=1}^n r_i t_i l_i }  \bigotimes_{i|s_i=1} |b_1+b_i+(1-c_i) t \rangle \right]
\end{equation}
Note that $r_i \in \{0,2\}$, so that we could write this as
\begin{equation}
{1 \over \sqrt{2}} \left[ \bigotimes_{i|s_i=1} |b_1+b_i+c_i t \rangle+(-1)^{\sum_{i=1}^n p_i t_i l_i }  \bigotimes_{i|s_i=1} |b_1+b_i+(1-c_i) t \rangle \right]
\end{equation}
where $p_i =r_i/2$.  Since $b_i$ and $c_i$ are given uniformly, we can rewrite this state as
\begin{equation}
{1 \over \sqrt{2}} \left[ |v\rangle + (-1)^{\sum_{i=1}^n p_i t_i l_i } |v+\bar{t}\rangle \right]
\end{equation}
where $v$ is a uniformly random element of ${\mathbb Z}_2^{ns}$, $s$ is the number of $s_i$ which equal $1$, and $\bar{t}$ is the vector in ${\mathbb Z}_2^{ns}$ which consists of $s$ repetitions of $t$.  We can further simplify this by applying double-controlled nots bitwise to this expression starting with the first $n$ elements acting on the second, the the second on the third, etc.  This then reduce the first $s$ ${\mathbb Z}_2^n$s to random vectors and the final vector will be
\begin{equation}
{1 \over \sqrt{2}} \left[ |w\rangle + (-1)^{\sum_{i=1}^n p_i t_i l_i } |w+t\rangle \right] \label{eq:ps}
\end{equation}
where $w$ is a uniformly random element of ${\mathbb Z}_2^n$.

Now we are in a similar situation to where we started, however instead of the phase being a power of $\omega$ the phase is a power of $-1$: in other words we have a procedure for producing from $n+2$ copies of states like that in Eq.~(\ref{eq:bsup}) a state of a similar form but with the phase doubled.  It therefore follows that if we apply this procedure to $n+2$ of the above states, we will double the phase once more.  Thus by using $(n+2)^2$ copes of the states in Eq.~(\ref{eq:bsup}) we can produce another doubling of the phase and therefore the state
\begin{equation}
{1 \over \sqrt{2}} \left[ |x\rangle + (-1)^{\sum_{i=1}^n 2p_i t_i l_i } |x+t\rangle \right] ={1 \over \sqrt{2}} \left[ |x\rangle +|x+t\rangle \right]
\end{equation}
This, however, is just the state which serves as input to Simon's algorithm for determining a hidden subgroup in ${\mathbb Z}_2^n$ where the hidden subgroup is $\{0,t\}$.  Thus by using $O(n)$ copies of this state we can use Simon's algorithm to determine $t$.  In total we have used $O(n^3)$ queries to determine the $t$.

\subsection{Determining the Parity of $l$}

Having determined $t$, we now show how to determine the parity of $l$.  Since we know $t$, we can now query the function $f$ over different superpositions where the $t_i=1$ and a fixed (irrelevant) bit string in the locations where $t_i=0$.  After doing this we will produce the following state over the locations where $t_i=1$:
\begin{equation}
{1 \over \sqrt{2}} \left( |b\rangle + \omega^{\sum_{i=1}^n -(-1)^{b_i} {\mu_i l_i}} |\bar{b}\rangle \right)
\end{equation}
If we now run the algorithm described in the previous subsection on $n+2$ copies of this state, we will obtain the state
\begin{equation}
{1 \over \sqrt{2}} \left( |v\rangle+ (-1)^{p_i l_i } |\bar{v}\rangle \right)
\end{equation}
where the $p_i$ are uniformly random elements of ${\mathbb Z}_2$ which we know.  Now suppose that we perform a measurement of such a state in the basis $|\psi_{x,\pm}\rangle={1 \over \sqrt{2}} \left( |x\rangle \pm |\bar{x}\rangle \right)$.  Then, depending on whether you get a $|\psi_{x,+}\rangle$ or $|\psi_{x,-}\rangle$, we know that the $l_i$ are solutions to
\begin{equation}
\sum_i p_i l_i = c {~\rm mod}~2
\end{equation}
where $c \in \{0,1\}$.  Thus by running this procedure $n$ times we will obtain a set of equations for the parity of the $l_i$ which, with high probability are linearly independent and can be obtained efficiently using Gaussian elimination.  Thus we have can determine the parity of the $l_i$ using $O(n^2)$ queries to the hidden subgroup oracle. 

\subsection{Determining $l$}

The final stage of the algorithm is determining $l$ beyond the parity of the $l_i$.  Note that having determined $t$ and the parity of the $l_i$ is equivalent to the hidden subgroup conjugacy problem for ${\mathcal D}_4^n$.  Suppose that we wish to determine the value of a particular $l_i$ and we know that parity of $l_i$ is such that $l_i \in \{0,2\}$.  When we are querying the function in the standard setup for the hidden subgroup problem we also compute an extra function in an ancilla register on the particular $i$th element of ${\mathcal D}_4^n$.  In particular if we calculate the function which maps the the element $r^t s^k$ to $k+2t~{\rm mod}~2$, then this function will be constant on the same subgroup as before if $l_i=0$.  If, however $l_i=2$, the subgroup will no longer be constant on this subgroup.  Thus by running, either of the procedures in the previous two subsections, we can determine which of these two cases hold and therefore whether $l_i=0$ or $l_i=2$.  A similar approach can be taken for $l_i=1$ versus $l_i=3$.  Further each of these can be carried out for the different $i$th terms in the hidden subgroup problem, and therefore $l_i$ can be efficiently computed for all $i$ using this procedure.

\section{Conclusion}

We have shown that the Gaussian elimination in Simon's algorithm can be interpreted in terms of a Clebsch-Gordan transform.  This led us to consider the hidden involution problem for ${\mathcal D}_4^n$.   Using Clebsch-Gordan transforms over ${\mathcal D}_4^n$, along with measurements on the multiplicity space of this transform, an efficient quantum algorithm for the hidden involution problem over ${\mathcal D}_4^n$ was derived.  An important open question is, of course, whether the approach of using Clebsch-Gordan transforms can be used to obtain efficient quantum algorithms for the diheral group or the symmetric group.  While we do not know the answer to this question, we do note that our algorithm uses the Clebsch-Gordan transform in a recursive fashion (first to filter the phase to powers of $(-1)$ and then to filter this phase to $1$.)  While this procedure appears to be a particular to dihedral group of order eight, it is important to investigate this approach for dihedral groups which are higher powers of two.  

\section{Acknowledgements}

We thank Andrew Childs for useful discussions of this work.  This work was supported by ARO/NSA quantum algorithms grant number W911NSF-06-1-0379 and NSF grant number 0523359 and NSF grant number 0621621.

\bibliographystyle{hunsrt}
\bibliography{../../../bigref/newbigref}

\end{document}